\documentclass[aps,prd,twocolumn,showpacs]{revtex4}
\usepackage{amsmath}
\usepackage{amsfonts}
\usepackage{graphicx}

\begin{document}

%% Title Page Stuff
  \title{$Z \rightarrow b\bar{b}b\bar{b}$ in the light gluino and light sbottom scenario}
  \author{Rahul Malhotra }
  \email[Email: ]{rahul@math.utexas.edu}
  \author{Duane A. Dicus}
  \affiliation{Center for Particle Physics, University of Texas, Austin, Texas, 78712, U.S.A.}
  \date{5 March, 2003}
  
  \begin{abstract}
The light gluino $(12 \sim 16$ GeV$)$ and light sbottom $(2 \sim 6$ GeV$)$ scenario 
has been used to explain the apparent overproduction of $b$-quarks at the Tevatron. 
This scenario also predicts the decay $Z \rightarrow b\bar{b}\tilde{g}\tilde{g}$ where the 
gluinos subsequently decay into $b$-quarks and sbottoms. We show that this can 
contribute to $\Gamma_{4b} = \Gamma(Z \rightarrow b\bar{b}b\bar{b})$ since most of the sbottoms and $b$-quarks 
arising from gluino decay have a small angular separation. We find that 
while no excess in $\Gamma_{4b}$ is observable due to 
large uncertainties in experimental measurements, the ratio  
$\Gamma(Z \rightarrow b\bar{b}\tilde{g}\tilde{g})/\Gamma(Z \rightarrow b\bar{b}b\bar{b})$ 
can be large due to sensitivity to $b$-quark mass, the sbottom mixing angle and the
gluino mass. We calculate it to be in the range $0.05 - 0.41$
inclusive of the entire parameter space.
  \end{abstract}
  \preprint{UTEXAS-HEP-156}
  \maketitle

%% INTRODUCTION
  \section{Introduction}
The Standard Model (SM) has been successful in explaining a host
of experimental observations on electroweak and QCD phenomenon at
LEP as well as hadron colliders. But it is generally believed that
the SM is an effective theory valid at the electroweak energy
scale with some new physics lying beyond it. Among the leading
candidates for a theory beyond the SM is the Minimal
Supersymmetric Standard Model (MSSM) \cite{haber} which has been
extensively studied in the past few decades.

The MSSM predicts the existence of SUSY partners of quarks, gluons
and other known particles of the Standard Model. These so-called
``sparticles'' have not been observed which has led to speculation
that they might be too heavy to have observable production rates
at present collider energies. However, it has been suggested in
\cite{lightsb} that a light sbottom ($\tilde{b}_1$) with a mass of
$\mathcal{O}(5$ GeV$)$ is not ruled out by electroweak precision
data if its coupling to the $Z$-boson is tuned to be small in the
MSSM. Recently Berger {\it et al} \cite{berger1} have also proposed a light sbottom
and light gluino (LSLG) model to explain the long-standing puzzle of
overproduction of $b$-quarks at the Tevatron \cite{excessb}.
In this model, gluinos of mass $12 - 16$ GeV are produced in pairs in $p\bar{p}$ collisions 
and decay almost immediately into a sbottom ($2 - 6$ GeV) and a $b$-quark each. 
The sbottom manages to evade direct detection via $R$-parity violating decays 
into soft jets of light quarks around the cone of the $b$-jet. 
This mechanism is shown to successfully fit 
the $b$-quark transverse momentum distribution at NLO level. Alternatively it has been 
suggested that using the latest $b$-quark
fragmentation functions reduces the discrepancy at the Tevatron \cite{cacciari}.
In this report however we will work in the LSLG scenario.

Recently, there has been some careful re-examination of
$Z$-pole precision data in this scenario. QCD corrections to the $Zb\bar{b}$
vertex with sbottom and gluino loops have been calculated \cite{cao}. They contribute negatively to $R_b$ 
and increase in magnitude with the mass of the other eigenstate of the sbottom ($\tilde{b}_2$). To maintain 
consistency with data, $\tilde{b}_2$ must be lighter than 125 (195) GeV at $2\sigma$ ($3\sigma$) level. 
An extension of that analysis to the entire range of electroweak precision data 
finds that $\tilde{b}_2$ must be lighter than $180$ GeV at
the $5\sigma$ level \cite{cho}. A $\tilde{b}_2$ in such a mass range would
have been produced in association with a $\tilde{b}_1$ at LEPII
energies (upto $209$ GeV) via the couplings $Z \tilde{b}_1 \bar{\tilde{b}}_2$ and
$Z \tilde{b}_2 \bar{\tilde{b}}_1$.  Since such a $\tilde{b}_2$ has
not been observed it would seem that LEP data disfavors the LSLG scenario. However, 
these constraints can be relaxed because (i) a subsequent study of the decay 
$Z \rightarrow \tilde{b}_1 \bar{b} \tilde{g} + \bar{\tilde{b}}_1 b \tilde{g}$ has shown 
that it can contribute positively to $R_b$ \cite{cheung1} and therefore should overcome at least some of 
the negative loop effects, (ii) a heavier $\tilde{b}_2$
($\gtrsim 200$ GeV) might be allowed
if large $CP$-violating phases are present in the model \cite{baek} and (iii) experimental searches for SUSY
particles are heavily model-dependent and, to our knowledge, 
an exhaustive search of LEPII data for a $\tilde{b}_2$ in
this particular scenario has not been done.

In addition to $Z$-precision data, production of $b\bar{b}$-pairs
at the $Z$-pole via gluon splitting has been re-examined recently by Cheung and Keung \cite{cheung}. 
They calculate the contribution of $Z \rightarrow q\bar{q}\tilde{g}\tilde{g}$ to the
process $Z \rightarrow q\bar{q}g^{*} \rightarrow b\bar{b}$ (q =
u,d,c,s,b) in the massless $q$ approximation and find that the
former is only around $4 - 15\%$ of the latter. They do not go
further and consider the ratio
 $\Gamma(Z \rightarrow b\bar{b}\tilde{g}\tilde{g})/\Gamma(Z \rightarrow b\bar{b}b\bar{b})$
 as they expect it to be very similar. Like them we note that the process $Z
\rightarrow b\bar{b}\tilde{g}\tilde{g}$ can contribute to $\Gamma (Z
\rightarrow b\bar{b}b\bar{b})$ via the final states
$b\bar{b}b\bar{b}\tilde{b}_1\bar{\tilde{b}}_1$,
$b\bar{b}bb\bar{\tilde{b}}_1\bar{\tilde{b}}_1$ and
$b\bar{b}\bar{b}\bar{b}\tilde{b}_1\tilde{b}_1$ which have four
$b$-quarks and two sbottoms arising from the gluino decays
$\tilde{g} \rightarrow b\bar{\tilde{b}}_1/\bar{b}\tilde{b}_1$. This can happen if the $b$-quark and sbottom 
arising from gluino decay prefer a small angular separation, so that a typical event looks 
like $Z \rightarrow b\bar{b}b\bar{b}$. But $Z \rightarrow
b\bar{b}\tilde{g}\tilde{g}$ can arise not only from the gluon
splitting diagrams in Fig. \ref{fig1}(b) but also from ``sbottom
 splitting'' diagrams in Fig. \ref{fig1}(c). We show that the net SUSY process should indeed 
contribute to $Z \rightarrow b\bar{b}b\bar{b}$ and the latter diagrams significantly
 enhance the width for this process.  We also calculate
 $\Gamma (Z \rightarrow b\bar{b}b\bar{b})$ to
leading order in the SM over a range of $b$-quark masses. The
final result is a wide theoretical range for the ratio $\Gamma(Z
\rightarrow b\bar{b}\tilde{g}\tilde{g})/ \Gamma(Z \rightarrow
b\bar{b}b\bar{b})$ of $5\%$ to $41\%$ inclusive of the entire
parameter space.

%% CALCULATIONS
  \section{Calculations}

\begin{figure}[h]
\includegraphics{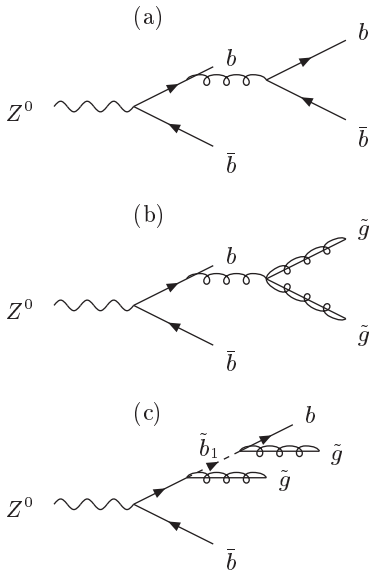}
\caption{\small \label{fig1}
Feynman diagrams contributing to (a) $Z
\rightarrow b\bar{b}b\bar{b}$ and (b),(c) $Z \rightarrow
b\bar{b}\tilde{g} \tilde{g}$. Diagrams with
gluon/gluino emission off the $\bar{b}$-leg and the crossing of
identical particles are not shown.
}
\end{figure}

The tree level diagrams for evaluating $\Gamma_{4b} = \left.
\Gamma (Z \rightarrow b\bar{b}b\bar{b}) \right|_{SM}$ and
$\Gamma_{b\tilde{g}} = \Gamma (Z \rightarrow
b\bar{b}\tilde{g}\tilde{g})$ are shown in Fig. \ref{fig1}. 

Feynman rules for the MSSM given by Rosiek \cite{rosiek} are used to evaluate
these diagrams. Their formalism allows us to write the lighter
sbottom mass eigenstate as a superposition $\tilde{b}_{1} =
\sin\theta_{b}$ $\tilde{b}_{L}$ $+$ $\cos\theta_{b}$
$\tilde{b}_{R}$ of the left and right-handed states where
$\theta_{b}$ is the sbottom mixing angle. This angle appears in
the coupling:
$$Z\tilde{b}_1\bar{\tilde{b}}_1 \propto (\frac{1}{2}
{\sin}^2\theta_b - \frac{1}{3} {\sin}^2\theta_W)$$ where
$\theta_W$ is the Weinberg angle. However since electroweak data
excludes the process $Z \rightarrow \tilde{b}_1\bar{\tilde{b}}_1$
to a high precision, the mixing angle must be fine-tuned to make
the coupling small i.e., $s_b^2 = \frac{2}{3}{\sin}^2\theta_W$,
$|s_b| \approx 0.38$ \cite{lightsb} where the short-hand notation
$s_b \equiv \sin\theta_b$ is used. We vary $|s_b|$ in the
narrow range $0.30 - 0.45$.

At constant scale we find that the diagrams in Fig. \ref{fig1}(c) enhance
the width for $Z \rightarrow b\bar{b}\tilde{g}\tilde{g}$ by $
10-60\%$. The lower limit is obtained for $s_b = -0.30$, $m_{\tilde{g}} = 12$ GeV and 
the upper for $s_b = +0.45$, $m_{\tilde{g}} = 16$ GeV. Default values of $m_b = 4.5$ GeV, 
$m_{\tilde{b}_1} = 4$ GeV are used in this analysis and variation of these within $m_b = 4 - 5.25$ GeV and 
$m_{\tilde{b}_1} = 2 - 6$ GeV has little effect. We also verify that as $m_{\tilde{b}_1}$ becomes large, the contribution 
of Fig. \ref{fig1}(c) diminishes, and vanishes in the limit $m_{\tilde{b}_1} \rightarrow \infty$. 
We therefore choose the invariant mass of the two gluinos, $m_{\tilde{g}\tilde{g}}$, as
the running scale $Q$ since the diagrams in Fig. \ref{fig1}(b) are still
dominant. Using a different invariant mass such as $Q = m_{b\tilde{g}}$ 
only changes $\Gamma_{b\tilde{g}}$ by $2-3\%$. 

In calculating $\Gamma_{4b}$ the $b\bar{b}$-pair produced by gluon
splitting cannot be isolated due to interference terms between
crossed diagrams in Fig. \ref{fig1}(a), making the off-shellness of the
virtual gluon indeterminate. This is in contrast to the gluon
splitting processes $Z \rightarrow q\bar{q}g^{*} \rightarrow
b\bar{b}$ and $Z \rightarrow b\bar{b}g^{*} \rightarrow q\bar{q}$,
$q \neq b$, where the secondary production of $b$-quarks does not
interfere with primary production at leading order \cite{kniehl}.
Therefore to calculate $\Gamma_{4b}$ we first find the ratio
$\Gamma_{4b}$/$\Gamma (Z \rightarrow q\bar{q}g^{*} \rightarrow
b\bar{b})$ at constant $Q$-scale, summing the denominator over $q
= u,d,s,c,b$ in the massless $q$ approximation. Then $\Gamma_{4b}$
is evaluated over a running $Q$-scale as follows:

\begin{equation}\label{eqn1}
\Gamma_{4b} = \underbrace{\frac{\Gamma_{4b}}{\Gamma (Z \rightarrow q\bar{q}g^{*} \rightarrow b\bar{b})}}_{Q = const.}
 \times \overbrace{\Gamma (Z \rightarrow q\bar{q}g^{*} \rightarrow b\bar{b})}^{Q = m_{b\bar{b}}}
\end{equation}

The gluon-splitting process has been studied up to next-to-leading logarithm (NLL) level in the past and 
is known to be sensitive to the $b$-quark mass. Values ranging from $m_b = 4.25$ GeV \cite{cheung} to variation 
between the pole mass and the $B$-meson mass i.e. $m_b = 4.75 - 5.25$ GeV have been used \cite{seymour1,seymour2}. 
DELPHI and OPAL \cite{delphi,opal} have measured $\Gamma_{4b}$ and they use $m_b = 4.5 - 5.25$ GeV in their 
analysis. Clearly, in the absence of a higher order calculation there is no choice but to vary $m_b$ over a wide range. 
We vary it from the lower limit of the
$\overline{MS}$ value $\bar{m}_b(\bar{m}_b) = 4$ GeV to the
$B$-meson mass $m_B = 5.25$ GeV \cite{PDG}. 

The canonical strong coupling value $\alpha_{S}(M_Z) = 0.118$ is used as changes to the running of 
$\alpha_S$ have been shown to be small \cite{berger1,alpha_s} even in light of the various new effects 
on the hadronic width of the $Z$ in this scenario \cite{lightsb,cao,cho,baek,cheung1,cheung,luo}.

Results are given in terms of the ratios:
$$R_{4b} \equiv \left. \frac{\Gamma_{4b}}{\Gamma(Z \rightarrow hadrons)} \right|_{SM},\quad R_{b\tilde{g}}
\equiv \frac{\Gamma_{b\tilde{g}}}{\Gamma(Z \rightarrow hadrons)}$$
The total hadronic width of the $Z$ is taken to be $\Gamma(Z \rightarrow hadrons) = 1.744$ GeV \cite{PDG}.

%% RESULTS
\section{Results}

Using eqn. (\ref{eqn1}) we numerically calculate $R_{4b} = (6.06 - 3.04)
\times 10^{-4}$ for $m_b = 4 - 5.25$ GeV. Measurements of the same by DELPHI 
and OPAL yield $(6.0\pm 1.9\pm 1.4) \times 10^{-4}$ and $(3.6\pm 1.7\pm 2.7)
\times 10^{-4}$ respectively and the uncorrelated average is given
by the Particle Data Group to be $R^{exp}_{4b} = (5.2\pm 1.9) \times 10^{-4}$
\cite{PDG}. Our entire calculated range is within $1.2\sigma$ of
the experimental average. Therefore, the lack of experimental precision does 
not allow us to fix the $b$-quark mass any further. The central value of $R^{exp}_{4b}$ is
obtained for $m_b \sim 4.3$ GeV which agrees well with $m_b =
4.25$ GeV used in \cite{cheung} to fit the full gluon-splitting process.
\begin{figure}[!]
\includegraphics{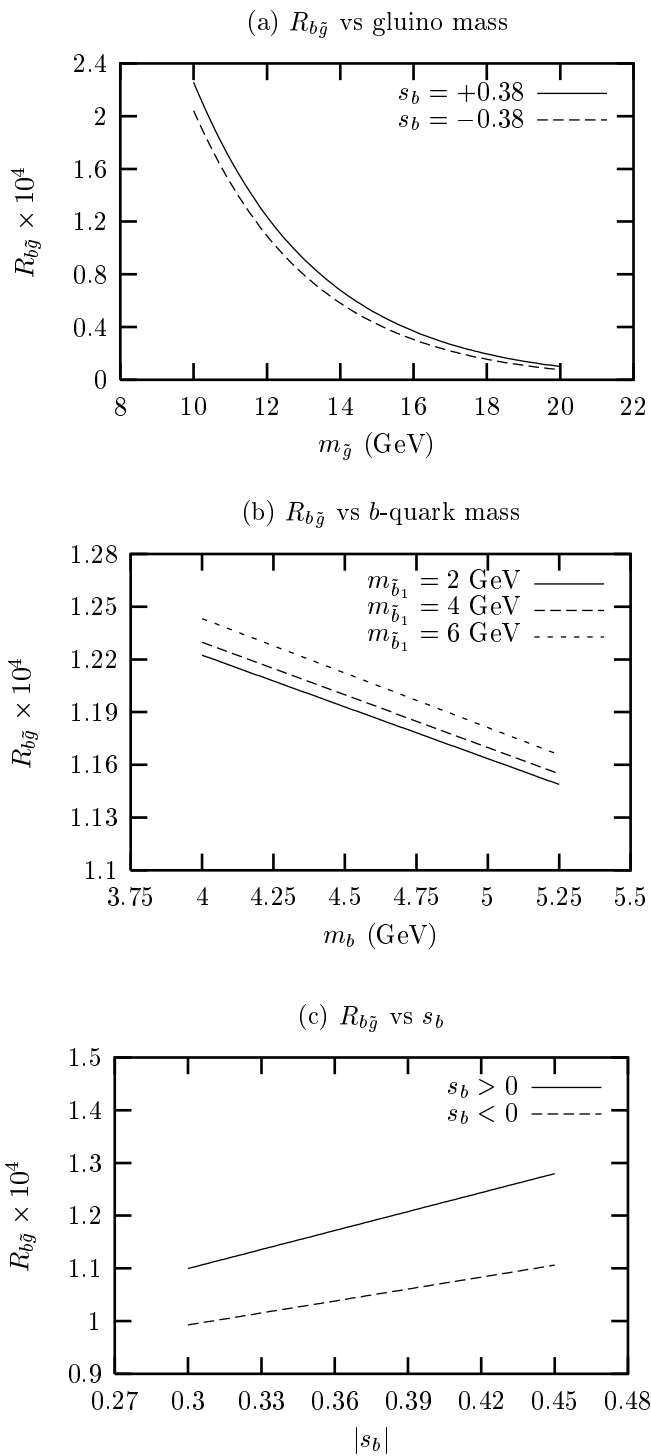}
\caption{\small \label{fig2}
Dependence of $R_{b\tilde{g}}$ on various parameters. If unspecified, 
default values of $m_{\tilde{g}} = 12$ GeV, $m_b = 4.5$ GeV, $m_{\tilde{b}_1} = 4$ GeV 
and $s_b = +0.38$ are used.
}
\end{figure}
\begin{figure}[!]
\includegraphics{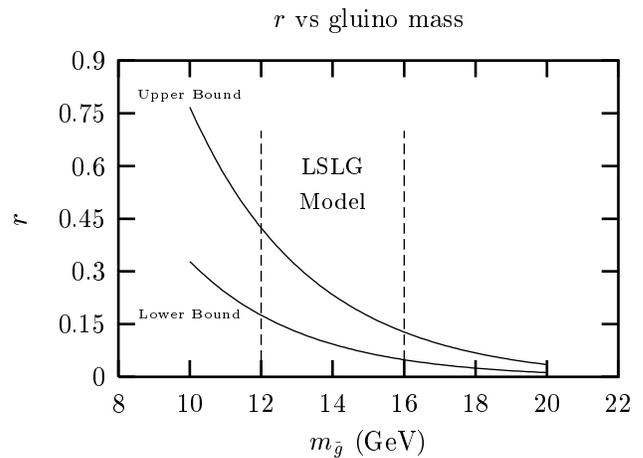}
\caption{\small \label{fig3}
$r = R_{b\tilde{g}}/R_{4b}$ dependence on gluino mass.
Upper (lower) bound curves are obtained for $m_b = 5.25$ ($4$)
GeV, $m_{\tilde{b}_1} = 6$ ($2$) GeV and $s_b = +0.45$ ($-0.30$).
}
\end{figure}
In a similar fashion, $R_{b\tilde{g}}$ is very sensitive to the
gluino mass $m_{\tilde{g}}$ showing a decrease by nearly a factor
of 3.5 as $m_{\tilde{g}}$ varies from $12$ to $16$ GeV [Fig. \ref{fig2}(a)]. 
On the other hand, variations in $b$-quark mass and the sbottom mass ($2 - 6$ GeV) have very little
effect ($\sim 5\%$) [Fig. \ref{fig2}(b)].
The effect of varying $s_b$ within the range $|s_b| = 0.30 - 0.45$ is shown in
Fig. \ref{fig2}(c) for $m_{\tilde{g}} = 12$ GeV. The variation is $\gtrsim 30\%$, increasing 
with gluino mass. $R_{b\tilde{g}}$ is lower for negative values of $s_b$
but increases with $|s_b|$ due to constructive interference with
the gluon-splitting diagrams. Including all parameters, we find
$R_{b\tilde{g}} = (0.25 - 1.33) \times 10^{-4}$.

The total $R_{4b} + R_{b\tilde{g}}$ equals $(3.3 - 7.3)
\times 10^{-4}$ for the entire parameter space which is still
within $1.2\sigma$ of the experimental value. However the ratio $r
= R_{b\tilde{g}}/R_{4b}$ can be quite large, varying from $5 -
41\%$ [Fig. \ref{fig3}]. Thus the SUSY process can be a significant 
fraction ($4 - 30\%$) of the total events if it cannot be distinguished from the SM 
$b\bar{b}b\bar{b}$ decay.

\begin{figure}[!]
\includegraphics{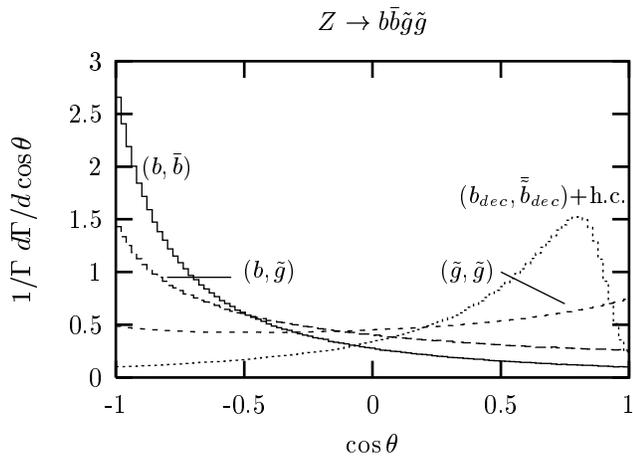}
\caption{\small \label{fig4}
Opening angles ($\cos \theta$) between $Z$-decay products. The prompt $b$-quarks are simply 
denoted by $b$ or $\bar{b}$ while $\tilde{b}_{dec}$ and $b_{dec}$ arise from gluino decay.
}
\end{figure}

In this context we now study the structure of $Z \rightarrow b\bar{b}\tilde{g}\tilde{g}$ events. The cumulative final state 
$q\bar{q}\tilde{g}\tilde{g}$ has been studied in $e^+ e^-$ annihilation at $\sqrt{s} = 189$ GeV \cite{cheung}, 
and our results are similar. Fig. \ref{fig4} shows the opening angle ($\cos \theta$) between final state $b$-quarks, 
gluinos and gluino decay products. We see that the prompt $b$-quarks tend to be 
well-separated from each other and also from the gluinos. We decay a 
gluino into a $b\bar{\tilde{b}}_1/\bar{b}\tilde{b}_1$ pair and find that 
the decay products are rather close to each other, with $\cos \theta$ 
peaking at around $0.8$. The two gluinos on the other hand appear to 
have a more or less uniform cosine distribution with a slight preference for 
smaller angles. Therefore most events primarily consist of four jets 
containing $b$-quarks, with at least three of them well-separated, and 
two of them having sbottoms in or quite close to them. If the sbottom hadronizes 
completely in the detector and deposits it's energy as a jet then it might be difficult to 
separate from the accompanying $b$-quark without a deliberate search. In that case 
the events would look just like $Z \rightarrow b\bar{b}b\bar{b}$. However, if the sbottom-jet lies 
near the periphery of the $b$-jet then it might widen it somewhat. A similar situation 
could arise if the sbottom undergoes $R$-parity violating decays into soft jets: 
$\bar{\tilde{b}}_1 \rightarrow u + s; c + d; c + s$ \cite{berger2}. Some of the $b$-jets in 
the $R^{exp}_{4b}$ sample might therefore be unusually wide and/or have a 
higher particle multiplicity. However, this can also happen due to radiation of hard gluons from 
$b$-quarks in $Z \rightarrow b\bar{b}b\bar{b}$, and might be difficult to distinguish. 
The remaining small fraction of events where the 
sbottom(s) are well-separated from the $b$-jet would compete 
with the background from $Z \rightarrow b\bar{b}b\bar{b} + j$ and 
$Z \rightarrow b\bar{b}b\bar{b} + 2j$ events and might require careful separation. Finally, if the sbottoms 
deposit only a fraction of their energy in the detector they 
might still be observed as missing energy. This energy is likely to be small though since 
sbottoms arise from gluino decay and tend to be rather soft in comparison to prompt 
$b$-quarks. Events with missing energy can also arise from $Z \rightarrow b\bar{b}b\bar{b}$ 
because of subsequent semileptonic decays $B \rightarrow X l \nu$. While a full detector simulation is beyond 
the scope of this report; given the experimental uncertainties and the absence of a 
deliberate search it is possible that 
sbottoms produced in this scenario escaped detection.
The SUSY events might therefore lie hidden in data for $Z \rightarrow b\bar{b}b\bar{b}$ and 
could be sufficient in number that a deliberate search might uncover them. 

%% CONCLUSIONS
\section{Conclusions}
We show that the light sbottom and light gluino scenario predicts the SUSY decay 
$Z \rightarrow b\bar{b}\tilde{g}\tilde{g}$. We also show that this can contribute to the SM process $Z \rightarrow b\bar{b}b\bar{b}$ 
because of gluino decay into pairs of proximate sbottoms and 
$b$-quarks. This scenario cannot 
be constrained by an excess in the measured rate $R^{exp}_{4b}$ due to large experimental and 
theoretical uncertainties. However, the ratio between the SUSY and SM decays 
is shown to be significant, from $5\%$ to $41\%$ inclusive of the entire 
parameter space. A good fraction of events in the $R^{exp}_{4b}$ sample might 
therefore arise from the SUSY process and contain sbottom signatures hidden in and around $b$-jets. 
We suggest a model-dependent search for sbottoms in existing LEP data in order to constrain this scenario.

%% ACKNOWLEDGEMENTS
\section{Acknowledgements}

This work was supported in part by the United States Department of
Energy under Contract No. DE-FG03-93ER40757.

\end{document}